\documentclass[aps,prx,amsmath,amssymb,twocolumn,showpacs,superscriptaddress,final,floatfix.longbibliography]{revtex4-2}
\usepackage{natbib}
\usepackage{graphicx}	
\usepackage{color}	
\usepackage[normalem]{ulem}
\usepackage{hyperref}
\hypersetup{
  colorlinks,
  citecolor=blue,
  linkcolor=blue,
  urlcolor=blue}

\newcommand{\RFA}{RbFe$_2$As$_2$}
\newcommand{\KFA}{KFe$_2$As$_2$}
\newcommand{\CFA}{CsFe$_2$As$_2$}
\newcommand{\BFA}{BaFe$_2$As$_2$}
\newcommand{\SRO}{Sr$_2$RuO$_4$}

\begin{document}

\title{Observation of Two Cascading Screening Processes in an Iron-based Superconductor}


\author{Ming-Hua Chang}
\email{These authors contributed equally.}
\affiliation{Department of Physics, The Pennsylvania State University, University Park, Pennsylvania, 16802 USA}
\author{Steffen Backes}
\email{These authors contributed equally.}
\affiliation{RIKEN iTHEMS,  Wako, Saitama 351-0198, Japan; Center for Emergent Matter Science, RIKEN, Wako, Saitama 351-0198, Japan}
\author{Donghui Lu}
\affiliation{Stanford Synchrotron Radiation Lightsource, SLAC National Accelerator Laboratory, Menlo Park, California, 94025 USA}
\author{Nicolas Gauthier}
\affiliation{Institut National de la Recherche Scientifique – Energie Matériaux Télécommunications, Varennes, QC J3X 1S2 Canada}
\author{Makoto Hashimoto}
\affiliation{Stanford Synchrotron Radiation Lightsource, SLAC National Accelerator Laboratory, Menlo Park, California, 94025 USA}
\author{Guan-Yu Chen}
\affiliation{Center for Superconducting Physics and Materials, National Laboratory of Solid State Microstructures and Department of Physics, Nanjing University, Nanjing 210093, China}
\author{Hai-Hu Wen}
\affiliation{Center for Superconducting Physics and Materials, National Laboratory of Solid State Microstructures and Department of Physics, Nanjing University, Nanjing 210093, China}
\author{Sung-Kwan Mo}
\affiliation{Advanced Light Source, Lawrence Berkeley National Laboratory, Berkeley, California, 94720 USA}
\author{Zhi-Xun Shen}
\affiliation{Stanford Institute of Materials and Energy Sciences, SLAC National Accelerator Laboratory, Menlo Park, California, 94025 USA}
\affiliation{Department of Physics, Stanford University, Stanford, California, 94305 USA}
\affiliation{Geballe Laboratory for Advanced Materials, Department of Applied Physics, Stanford University, Stanford, California, 94305 USA}
\author{Roser Valent\'\i}
\email{valenti@itp.uni-frankfurt.de}
\affiliation{Institut f\"{u}r Theoretische Physik, Goethe-Universit\"{a}t Frankfurt, Max-von-Laue-Str.~1, 60438 Frankfurt am Main, Germany}
\author{Heike Pfau}
\email{heike.pfau@psu.edu}
\affiliation{Department of Physics, The Pennsylvania State University, University Park, Pennsylvania, 16802 USA}

\date{\today}


\begin{abstract}
Understanding how renormalized quasiparticles emerge in strongly correlated electron materials provides a challenge for both experiment and theory. It has been predicted that distinctive spin and orbital screening mechanisms drive this process in multiorbital materials with strong Coulomb and Hund’s interactions. Here, we provide the experimental evidence of both mechanisms from angle-resolved photoemission spectroscopy on {\RFA}. We observe that the emergence of low-energy Fe 3$d_{xy}$ quasiparticles below 90K coincides with spin screening. A second process changes the spectral weight at high energies up to room temperature. Supported by theoretical calculations we attribute it to orbital screening of Fe $3d$ atomic excitations. These two cascading screening processes drive the temperature evolution from a bad metal to a correlated Fermi liquid.
\end{abstract}

\maketitle


\section*{Introduction}

Metals with strong electronic correlations are characterized by renormalized quasiparticles 
with large effective masses. At high temperatures, coherent quasiparticle excitations at low energies are absent. Instead, high energy atomic excitations materialize as Hubbard bands and dominate the spectral function. Dispersive bands of coherent quasiparticles can emerge upon lowering the temperature due to screening of the atomic degrees of freedom.

Describing these screening processes in hallmark strongly correlated multi-orbital $d$ electron materials, such as iron-based superconductors (FeSCs) and ruthenates, remains challenging~\cite{dai_2015_review,dagotto_2013_review,Qazilbash_2009,Hayes_2016,Tokura_2000,Schaffer_2016,Mackenzie_2017_review}. In these systems, the local Coulomb interaction $U$ and a sizable Hund's rule coupling $J$ give rise to a multitude of unique 
correlation effects \cite{neupane_2009,Yi_2017,Medici_2014,Kugler_2019,Yi_2015,Huang_2022}. The notable sensitivity of their electronic properties on the Hund's coupling $J$ has led to their classification as so-called Hund's metals~\cite{Haule_2009,Medici2011,Medici2011PRL,Yin_2011,Georges_2013,Medici_2014,Backes_2015,Stadler2019,Villar2021,Crispino2023}. Starting from high temperatures, Hund's metals are predicted to separately screen first orbital and then spin degrees of freedom. They become Fermi liquids once both screening processes are complete \cite{Stadler_2019,deng_2019,horvat2019,Stadler_2021,suzuki_2023}. Here, we experimentally demonstrate that two distinct screening processes drive the emergence of long-lived, heavy quasiparticles in the spectral function of FeSCs.

\begin{figure*}
\includegraphics[width=\textwidth]{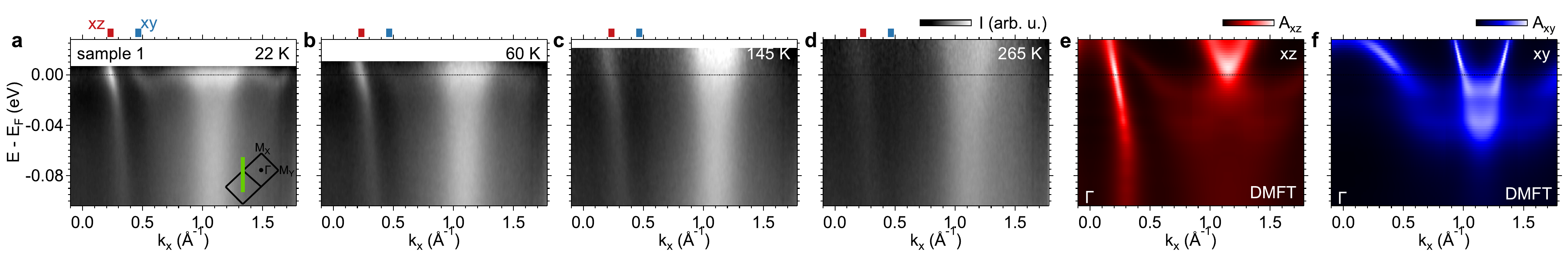}
\caption{
Temperature dependence of the low-energy electronic structure of \RFA. (a-d) ARPES spectra on sample 1 for selected temperatures. The green line in the inset sketches the momentum cut through the Brillouin zones. The dot indicates normal emission geometry. The spectra are divided by a Fermi-Dirac distribution and the intensity scale is the same for all spectra. The red and blue squares in (a-d) indicate the momentum integration range for the EDC analysis shown in Fig.~\ref{Fig:QP_spectral_weight}. (e,f): Spectral function calculated by DFT+DMFT at 96\,K and projected onto the $d_{xz}$ and $d_{xy}$ orbitals, respectively. The momentum direction and range is the same as for ARPES.
}
\label{Fig:spectra}
\end{figure*}
\begin{figure}
\includegraphics[width=\columnwidth]{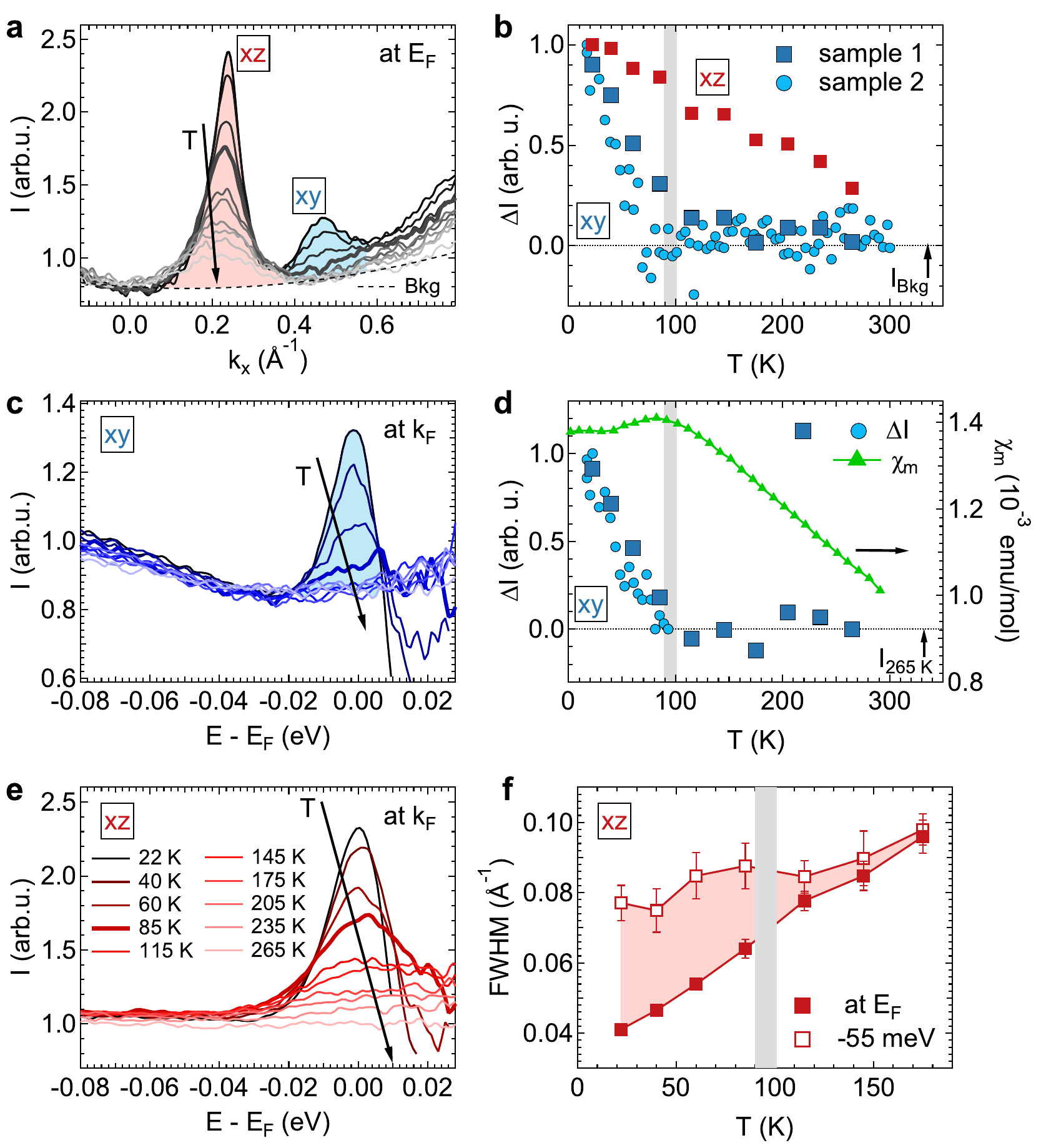}
\caption{
Quantitative quasiparticle spectral weight analysis. (a) MDCs integrated within $\pm 5$ meV around $E_F$. The dashed line shows a quadratic approximation of the high-temperature background within (-0.15,0.6)\textup{~\AA}$^{-1}$. Temperature values (dark to bright) are labeled in (e). (b) Change of the spectral weight obtained from integrating the MDCs in (a) within the momentum range marked by the shaded areas. $\Delta I = I - I_{\mathrm{Bkg}}$ is normalized to the value at 22 K. (c) EDCs at $k_F$ of the $d_{xy}$ hole band integrated within $(0.44,0.49)\,\mathrm{\AA}^{-1}$ as indicated by the blue box above the spectra in Fig.~\ref{Fig:spectra}. (d) Spectral weight obtained from the EDCs in (c) by integration within the energy window marked by the shaded area. $\Delta I = I - I_{\mathrm{265K}}$ is normalized to the value at 22 K. The $d_{xy}$ spectral weight response is compared to the magnetic susceptibility from Ref.~\onlinecite{Khim_2017}. (e) Same as in (c) but at $k_F$ of the $d_{xz}$ hole band and integrated within $(0.21,0.26)\,\mathrm{\AA}^{-1}$. (f) FWHM of the $d_{xz}$ band obtained from fits to the MDCs in (a). The plotted values are the average FWHM within $(-55\pm15)$\,meV and $(0\pm9)$\,meV. Error bars represent the standard deviation. The raw MDCs and EDCs for sample 2 are shown in Supplementary Fig.~S2 \cite{supplement}.
}
\label{Fig:QP_spectral_weight}
\end{figure}

We investigate \RFA, a hole-doped variant of the 122 parent compound \BFA. It is one of the most strongly correlated representatives of the Hund's metal FeSCs. Strong correlations were observed in low temperature thermodynamic and transport measurements, which found a large Sommerfeld coefficient of 127\,mJ/molK$^2$ and large effective masses up to $24 m_e$ \cite{Khim_2017,Zhang_2015,Eilers_2016}. 
Nominally 5.5 electrons occupy the five Fe $3d$ orbitals, which exhibit orbitally-differentiated correlation effects. The Fe $3d_{xy}$ orbital is the most correlated, followed by the degenerate $d_{xz}$ and $d_{yz}$ orbitals.
{\RFA} is a bad metal at high temperatures and displays an incoherent-coherent crossover when the temperature is lowered \cite{Wu_2016,Xiang_2016,Wiecki_2021}. Fermi liquid behavior sets in below 45\,K \cite{Xiang_2016}. 
Evidence for spin screening was obtained from magnetic susceptibility measurements that show signatures of local moments at high temperatures and of a Pauli spin susceptibility below 90\,K \cite{Khim_2017,Wu_2016}.

We use angle-resolved photoemission spectroscopy (ARPES) to follow the complete evolution of the Hund's metal spectral function in \RFA~across high and low energies and from high to low temperatures. 
We observe the sudden emergence of the $d_{xy}$ quasiparticle below 90\,K. 
It coincides with the peak in the magnetic susceptibility~\cite{Khim_2017,Wu_2016} that is related to spin screening. 
We identify an additional screening process that smoothly changes the $d_{xz}$ quasiparticle intensity as well as the intensity at high energies up to 6\,eV. Based on a combination of a simple 5-orbital atomic model and density functional theory + dynamical mean-field theory (DFT+DMFT), we connect the high-energy intensity changes to atomic multiplet excitations. We propose that orbital screening is responsible and extends beyond room temperature. Orbitals are screened at higher temperatures and spins at lower temperatures creating a cascade of successive screening processes.

\section*{Results and Discussion}

\subsection*{Low energies -- Quasiparticles}

We first discuss the orbital-dependent evolution of the quasiparticle spectral weight as a function of temperature at low energies around the Fermi energy $E_F$ between (-0.08,0.02)\,eV. Figure~\ref{Fig:spectra}(a-d) presents ARPES spectra for selected temperatures. Surface bands were suppressed by temperature cycling (Methods, Supplementary Fig.~S1,S2 \cite{supplement}). Data on a second, pristine sample show the same temperature dependence of the quasiparticle spectral weight \cite{supplement}.

$k_x$ and $k_y$ are equivalent momentum directions in the tetragonal crystal structure of \RFA. For ease of readability, we label the measured momentum direction as $k_x$ throughout this manuscript. Orbital degeneracy implies that our results for the $d_{xz}$ orbital also apply to the $d_{yz}$ orbital. 

Photoemission matrix elements for the experimental geometry favor emission from $d_{xz}$, $d_{xy}$, and $d_{z^2}$ orbitals (Supplementary Fig.~S5 \cite{supplement}) \cite{yi_2019_nematic,pfau2019detailed,li_2024_matrix}. We focus here on the two hole bands around the Brillouin center, which we identify as predominantly $d_{xz}$ and $d_{xy}$ character using DFT+DMFT calculations (Fig.~\ref{Fig:spectra}(e,f)). This assignment is in agreement with previous ARPES studies on the sister compounds \CFA~and \KFA~\cite{Richard_2018,Yoshida_2014,fang_2015}. DMFT also captures the experimentally observed dispersion renormalization of the hole bands at $E_F$ in \RFA~\cite{Chang_2024}. We present an analysis of the bands at the Brillouin zone corner in the Supplementary Fig.~S3\cite{supplement}.

The spectra in Fig.~\ref{Fig:spectra} indicate that the quasiparticle bands broaden substantially and the $d_{xy}$ band vanishes at high temperatures. We quantify this behavior with a detailed analysis of energy and momentum distribution curves (EDCs and MDCs) in Fig.~\ref{Fig:QP_spectral_weight}. Both the $d_{xz}$ and $d_{xy}$ quasiparticle peaks are well developed in the MDCs at low temperatures (Fig.~\ref{Fig:QP_spectral_weight}(a)). Their intensity decreases with increasing temperature and the $d_{xy}$ quasiparticle peak vanishes completely around 90\,K (Fig.~\ref{Fig:QP_spectral_weight}(b)). A clear $d_{xz}$ peak can be observed up 300\,K.

A complementary EDC analysis at $k_F$ confirms this finding. The integrated area below the $d_{xy}$ peak vanishes around 90\,K, beyond which the EDCs become temperature independent (Fig.~\ref{Fig:QP_spectral_weight}(c,d)). The peak intensity of the $d_{xz}$ orbital continues to decrease at high temperatures (Fig.~\ref{Fig:QP_spectral_weight}(e)). The integrated $d_{xz}$ peak area cannot be extracted for all temperatures since the peak broadens beyond the accessible energy range above $E_F$. This broadening mimics recent ARPES results on \SRO~\cite{hunter_2023}. The $d_{xz}$ width in the EDCs sharpens below 85\,K. The comparison of the full width at half maximum (FWHM) between $E_F$ and -55\,meV in Fig.~\ref{Fig:QP_spectral_weight}(f) illustrates the effect. Above 90\,K, the FWHM at both energies is almost identical. Below that temperature, the FWHM at -55\,meV saturates while it decreases more rapidly at $E_F$ (see also \cite{Chang_2024}). The qualitatively different behavior of the $d_{xz}$ and $d_{xy}$ orbital is a consequence of orbital differentiation in FeSCs. The $d_{xy}$ orbital is the most correlated one in \RFA~and in FeSCs in general \cite{Yin_2011,Medici_2014,diehl_2014,Backes_2015,Eilers_2016,Yi_2017}.

The appearance of the $d_{xy}$ quasiparticle at low temperatures was also observed in other FeSCs \cite{Yi_2013,Yi_2015,Pu_2016,Huang_2022}. In addition, we compare here the $d_{xy}$ ARPES intensity with magnetic susceptibility measurements (Fig.~\ref{Fig:QP_spectral_weight}(d)). The peak in $\chi_m$ coincides with the appearance of the $d_{xy}$ orbital. The temperature dependence of $\chi_m$ in \RFA~was interpreted as a crossover from large fluctuating moments with a Curie behavior at high temperatures to screened moments with an enhanced susceptibility and Pauli behavior at low temperatures \cite{Wu_2016,Hardy_2013}. The scaling between the magnetic susceptibility and the Knight shift deviates at the same crossover temperature \cite{Wu_2016}. It is reminiscent of the ubiquitous Knight shift anomaly in heavy fermion materials that is caused by spin screening \cite{jiang_2014}. Spin screening in \RFA~coincides with the coherence-incoherence crossover observed in resistivity \cite{Wu_2016} as well as signatures in the Hall effect \cite{Xiang_2016} and elastoresistance \cite{Wiecki_2021}. The crossover temperature scales along the (K,Rb,Cs)Fe$_2$As$_2$ series and the signatures are in line with theoretical predictions of spin screening in FeSC in general \cite{Haule_2009,Yin_2011}. To our knowledge, there is so far no microscopic theory that can account for this correlation.

The $d_{xy}$ quasiparticle spectral weight and the lifetime of the $d_{xz}$ orbital increase throughout the spin screening crossover in \RFA. They signify the emergence of long-lived quasiparticles. The Fermi surface in the bad metal regime at high temperatures is dominated by short-lived $d_{xz}$ excitations.

\subsection*{High energies -- Atomic excitations}

\begin{figure}
\includegraphics[width=\columnwidth]{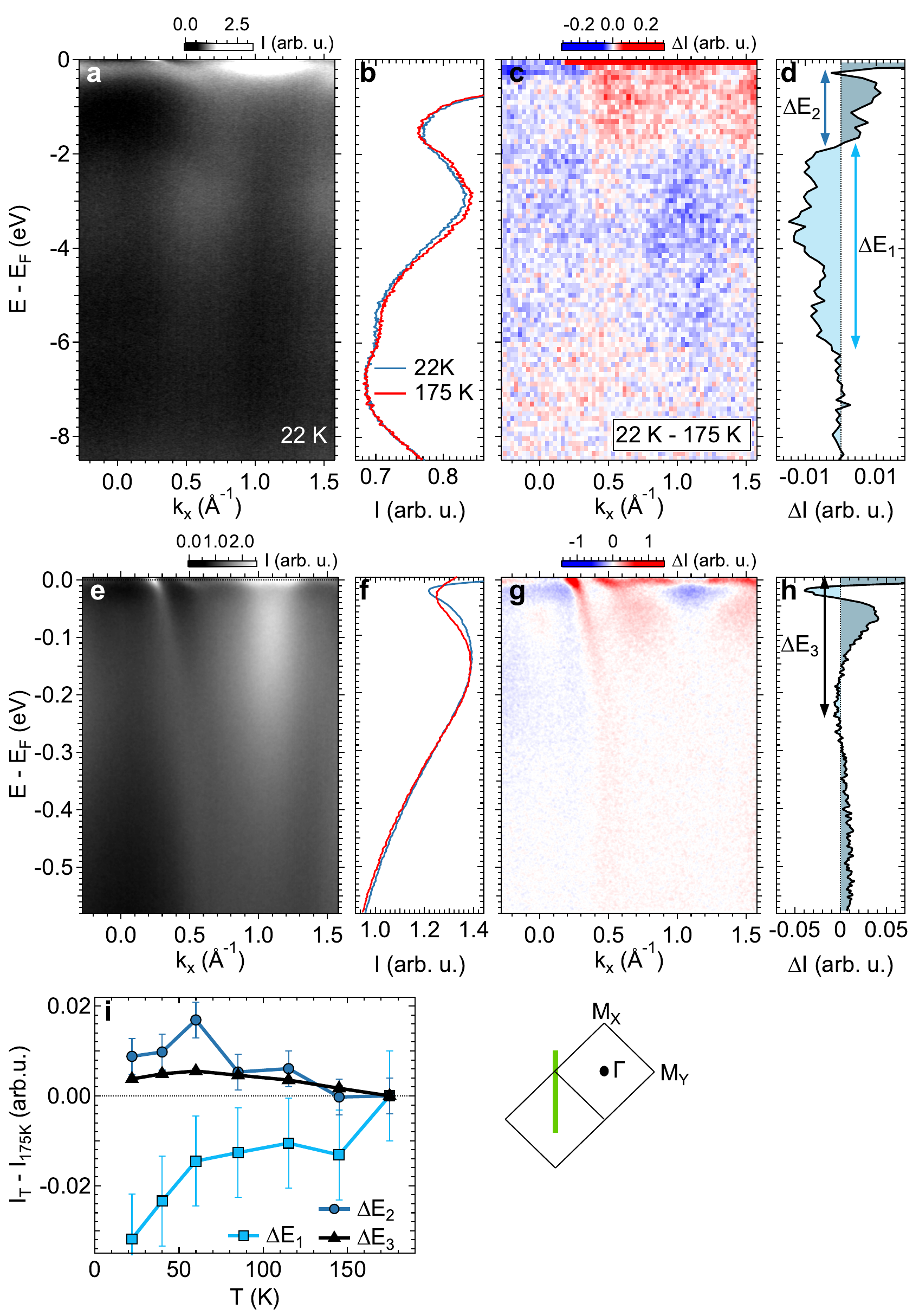}
\caption{
Temperature dependence of the high-energy electronic structure. (a) ARPES spectrum at 22\,K and up to 8.5\,eV. (b) EDCs integrated over the whole momentum range at 22\,K and 175\,K, respectively. (c) Intensity difference between spectra at 22\,K and 175\,K. (e-h): Same as (a-d) but with a zoom into the intermediate energy region up to 0.6\,eV. All spectra are are divided by a Fermi-Dirac distribution and are along the same momentum cut as in Fig.~\ref{Fig:spectra}, as indicated by the sketch of the Brillouin zones at the bottom. (i) Temperature dependence of the intensity for three different energy windows as defined in (d,h). The intensity at 175\,K is taken as reference. Error bars are determined on the basis of the standard deviation of the EDCs in (b).
}
\label{Fig:hubbard}
\end{figure}

\begin{figure*}
\includegraphics[width=\textwidth]{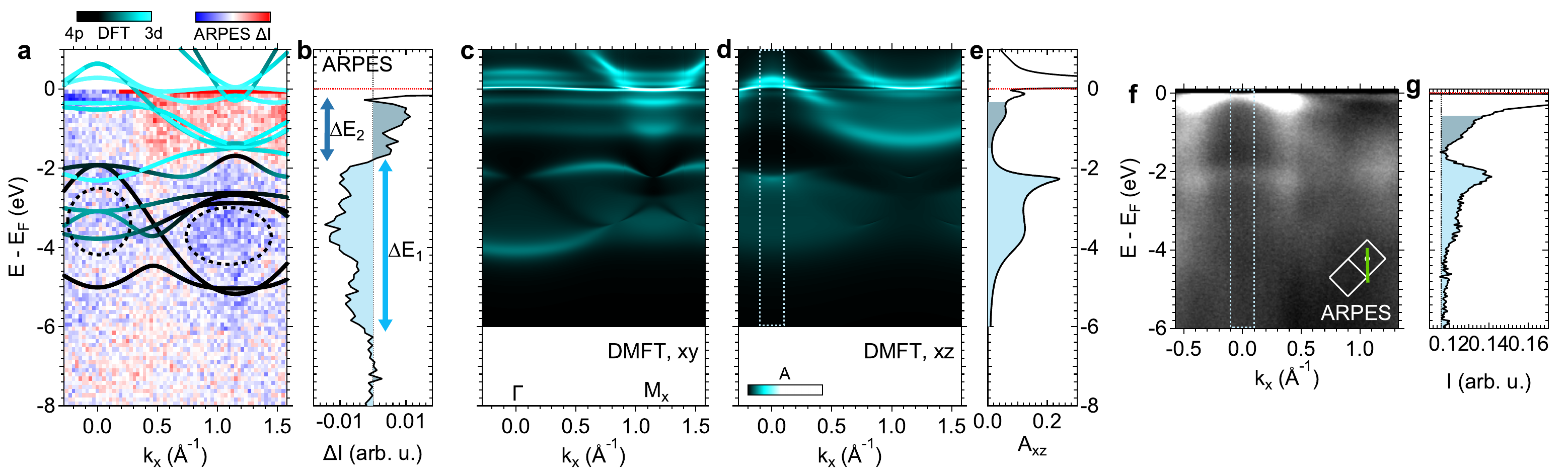}
\caption{
Comparison of ARPES with electronic structure calculations. (a) DFT bandstructure of \RFA. Dispersions are overall scaled by a factor of 1.15 to match the $4p$ band dispersion observed in ARPES (see Supplementary Fig.~S4 \cite{supplement}). The color indicates the contribution from Fe $3d$ and As $4p$ orbitals. DFT is plotted on top of the difference in ARPES intensity $\Delta I$ between 22\,K and 175\,K reproduced from Fig.~\ref{Fig:hubbard}(c). Dashed circles mark areas of small and large response in $\Delta I$ as discussed in the main text. (b) EDC for the difference spectrum in (a). (c,d): Spectral function calculated by DFT+DMFT and projected onto the $d_{xy}$ and $d_{xz}$ orbitals, respectively. (e) EDC from (d) integrated over the momentum range indicated by the box in (e). (f) ARPES spectrum of \RFA~taken in the first Brillioun zone. (g) EDC from (f) in the same momentum integration window as for the EDC in (e).
}
\label{Fig:hubbard_DMFT}
\end{figure*}

\begin{figure}
\includegraphics[width=\columnwidth]{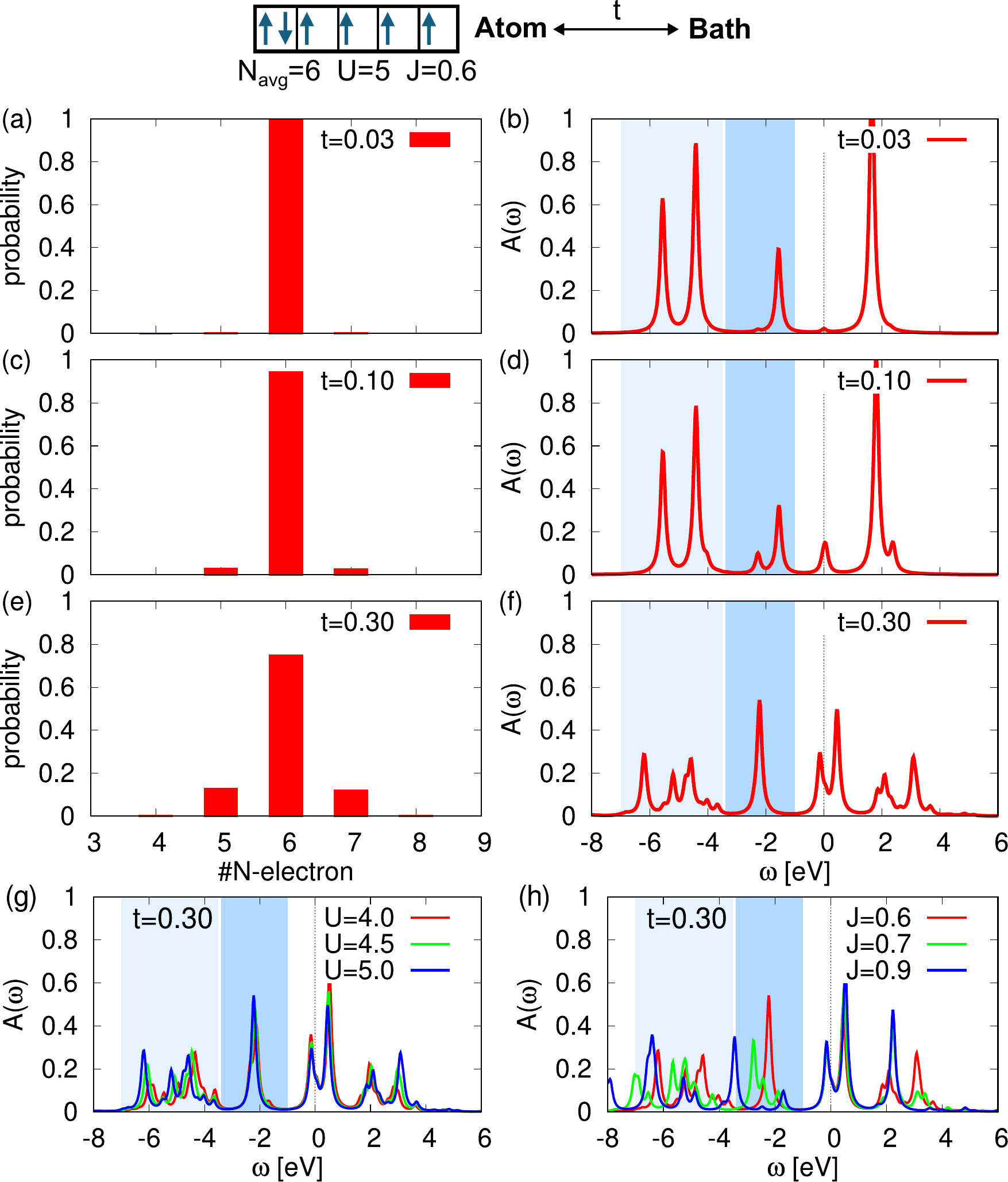}
\caption{
Simplified atomic model of 5 degenerate orbitals. (a) Probability of different occupation numbers for a coupling $t=0.03$ to the bath. We used a Kanamori interaction with $U=5$ and $J=0.6$, which are typical values of Coulomb interaction and Hund's coupling in Fe-based superconductors. (b) Corresponding spectral function. (c)-(f) Same as (a) and (b) but for different coupling $t$. (g) Spectral function for different values of the Coulomb interaction $U$ and at $J=0.6$. (h) Spectral function for different values of the Hund's rule coupling $J$ and at $U=5$. Both (g) and (h) are calculated at $t=0.3$. We mark electron removal states that depend on $U$ and $J$ with light background and those that only depend on $J$ with dark background.
}
\label{Fig:atomic_model}
\end{figure}

Spectral weight conservation dictates that the spectral function decreases at large positive or negative energies when it increases around $E_F$. We therefore present temperature-dependent ARPES up to 8.5\,eV in Fig.~\ref{Fig:hubbard} to probe the occupied states at large binding energies. 
 
The spectra show diffuse spectral weight beyond 0.3\,eV in contrast to the sharp quasiparticle bands around the Fermi level (Fig.~\ref{Fig:hubbard}(a,e)). The intensity changes with temperature in an energy and momentum dependent fashion, which is highlighted in difference plots between the spectra at 22\,K and 175\,K in Fig.~\ref{Fig:hubbard}(e) and (g). EDCs of the difference spectra (Fig.~\ref{Fig:hubbard}(d,h)) indicate three distinct energy regions $\Delta E_1=(6.1,1.8)$\,eV, $\Delta E_2 = (1.8,0.24)$\,eV and $\Delta E_3 = (0.24,0)$\,eV. The intensity increases with temperature in $\Delta E_1$ and it decreases in $\Delta E_2$ and $\Delta E_3$ (Fig.~\ref{Fig:hubbard}(i)). The intensity is temperature independent beyond 6.1\,eV. 

The integrated spectral weight within $\Delta E_3$ is predominantly due to photoemisison from the $d_{xz}$ orbital (Supplementary Fig.~S5 \cite{supplement}. The intensity is almost constant below 50\,K and it decreases at higher temperatures. We have discussed the changes of the quasiparticle bands close to $E_F$ in detail above.

To identify the origin of the temperature dependence at high binding energies, we compare our experimental results with DFT calculations in Fig.~\ref{Fig:hubbard_DMFT}(a). The As $4p$ bands are located within $\Delta E_1$, while the Fe $3d$ bands lie within $\Delta E_2$. As expected, DFT reveals $p$-$d$ orbital mixing. A change in orbital mixing due to thermal expansion is therefore an obvious interpretation of the different temperature evolution in $\Delta E_1$ and $\Delta E_2$. However, the following arguments render this scenario unlikely. 

1) When the lattice expands then electron overlap and hence $4p$-$3d$ mixing decreases. As a consequence, $3d$ ($4p$) spectral weight decreases (increases) in $\Delta E_1$ and increases (decreases) in $\Delta E_2$ (See Supplementary Note 1 and Supplementary Fig.~S6 \cite{supplement}). The sign of the intensity change can therefore only be observed in case we predominantly photoemit from As $4p$ bands. The precise photon-energy dependence of the photoemission cross section is rather complex. However, 60\,eV photons used here generally emphasize $3d$ spectral weight over $4p$ \cite{yeh_1985,Evtushinsky_2016,Watson_2017,Pfau_2021}. 

2) The in-plane length change between 22\,K and 175\,K is $\Delta L/L=5\times10^{-3}$ \cite{Wiecki_2021}. DFT predicts that the relative orbital contribution changes by the same amount (see Supplementary Fig.~S6) \cite{supplement}. But the ARPES intensity change $\Delta I/I\leq2.5\times10^{-2}$ is a factor of 5 larger. 

3) The intensity changes most strongly away from the As $4p$ bands and very little where $p$-$d$ overlap is largest (see dashed circles in Fig.~\ref{Fig:hubbard_DMFT}(a)).

In the following, we show that our temperature-dependent ARPES data can be interpreted as screening of atomic multiplet excitations. 
The simplified atomic model calculation in Fig.~\ref{Fig:atomic_model} illustrates the excitation spectrum of an atomic site with five degenerate, interacting orbitals. They are occupied on average by $N_\mathrm{avg}=6$ electrons. If isolated, the ground state of the atom is the high-spin configuration with $S=2$ and $N=6$. We expect three electron removal states and one addition state within an energy range on the order of the Coulomb interaction $U$ and the Hund's rule coupling $J$ (main peaks in Fig.~\ref{Fig:atomic_model}(b)).

We can model screening by coupling the atomic site to a discretized bath with five states through the hybridization amplitude $t$. The larger $t$, the more the electron number fluctuates due to charge exchange with the bath. We plot the probability of different occupation numbers for selected $t$ in Fig.~\ref{Fig:atomic_model}(a,c,e). The corresponding spectral functions are shown in Fig.~\ref{Fig:atomic_model}(b,d,f). 

Peaks emerge in the spectral function at the Fermi level and their intensity increases with $t$. The peaks correspond to the renormalized quasiparticle peaks in the full DMFT model. This effect may be associated with the increase of the ARPES intensity close to the Fermi level within $\Delta E_3$.

At high energies (shaded regions), several peaks appear in the atomic spectrum as function of $t$ and all shift in energy. The complexity is a consequence of the multi-orbital nature and various aspects of it have been studied previously \cite{Huang2014,Stadler_2019,boidi_2021,Stadler_2021,Sroda2023}. We can identify two groups of excitations from a detailed analysis of their $U$ and $J$ dependence (Fig.~\ref{Fig:atomic_model}(g,h)). The energy of states in the light shaded region ($-8$\,eV to $-4$\,eV) depend on both $U$ and $J$ as is typical for Hubbard bands. Their overall intensity decreases with $t$. The states in the dark shaded region ($-3$\,eV to $-1$\,eV) only depend on $J$ and were previously termed Hund bands \cite{Sroda2023}. In contrast to the Hubbard bands, their intensity varies non-monotonically with $t$. 

We can identify atomic excitations both in the DFT+DMFT and ARPES spectral function of {\RFA}. Dispersionless bands appear at approximately 0.3\,eV and 1\,eV in DFT+DMFT projected onto the $d_{xy}$ orbital (Fig.~\ref{Fig:hubbard_DMFT}(c)). All other orbitals possess similar weakly dispersing bands (Fig.~\ref{Fig:hubbard_DMFT}(d)). 
They are separated from the As $4p$ bands and do not appear in DFT calculations. Therefore, we interpret them as multiplet atomic excitations, not captured within a DFT-only approach. Our ARPES spectra show a corresponding peak (Fig.~\ref{Fig:hubbard_DMFT}(f)), which is most obvious as a shoulder at 1\,eV in the EDC at $\Gamma$ (Fig.~\ref{Fig:hubbard_DMFT}(g)). The EDC from DMFT (Fig.~\ref{Fig:hubbard_DMFT}(e)) matches the ARPES measurement. Local interactions are therefore well suited to describe \RFA. The dispersionless bands are reminiscent of Hubbard bands in FeSe \cite{Evtushinsky_2016,Watson_2017,Pfau_2021}. From the size of $U$ and $J$, we expect the multiplet to extend several eV below $E_F$ \cite{Chang_2024}. However, it is challenging to identify atomic excitations at larger energies due to the overlap with the As $4p$ bands.

A comparison between the ARPES data and the atomic model from Fig.~\ref{Fig:atomic_model} suggests that states within $\Delta E_1$ correspond to the Hubbard bands of the atomic model (light shaded region). Similarly, states within $\Delta E_2$ correspond to the Hund bands (dark shaded region). This distinction and their different temperature dependence is a consequence of the multi-orbital nature of \RFA. Understanding this complexity allows us to identify screening as the basic underlying mechanism that drives the temperature evolution of the high-energy ARPES intensity.

Recent temperature-dependent DMFT calculations of a three-orbital Hund-Hubbard model with sizable $J$ confirm this interpretation \cite{Stadler_2021}. The spectral function has the opposite temperature dependence around the low-energy atomic excitation than around those at high energies. The sign agrees with our experimental observation.


\subsection*{Discussion -- Spin and orbital screening}

\begin{figure}
\includegraphics[width=\columnwidth]{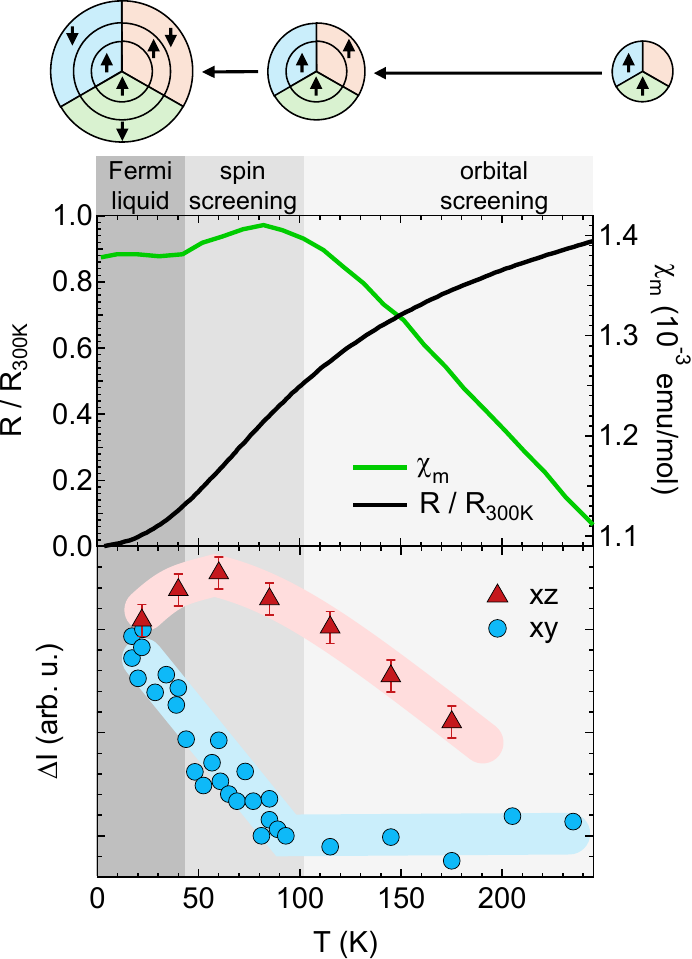}
\caption{
Cascade of screening processes. A comparison of the resistivity $R$ (from Ref.~\onlinecite{Wiecki_2021}), the magnetic susceptibility $\chi_m$ (from Ref.~\onlinecite{Khim_2017}) and the quasiparticle spectral weight change $\Delta I$ of the $d_{xy}$ and $d_{xz}$ orbitals (from Fig.~\ref{Fig:QP_spectral_weight}(d) and Fig.~\ref{Fig:hubbard}(i)). Top: Sketches illustrate orbital and spin screening in the example of three atomic orbitals (circular segments) occupied by two electrons (adapted from Ref.~\onlinecite{georges_2024}). Right: Hund's rule coupling favors a parallel spin alignment of both electrons in the atomic orbitals. Center: Orbital screening (second circle) of the atomic configuration leads to an orbital singlet. Hund's rule coupling favors the large spin state. Left: Spin screening (third circle) of the large moment then leads to the formation of a spin singlet. The Fermi liquid regime has a constant $\chi_m$ and a $T^2$ dependence of $R$. Spin screening is identified as the crossover between Pauli and Curie behavior in $\chi_m$. It coincides with a crossover in the resistivity (slope change). The $d_{xy}$ quasiparticle emerges once spin screening sets in. Orbital screening leads to a spectral weight transfer from the Hubbard bands to the $d_{xz}$ quasiparticle peak. Its intensity increases over the whole screening temperature range.
}
\label{Fig:screening}
\end{figure}

The temperature dependence of the high energy spectral weight (Fig.\,\ref{Fig:hubbard}(i)) does not change significantly across the spin screening temperature of 90\,K that we identified from the magnetic susceptibility. Simultaneously, the low-energy $d_{xz}$ quasiparticle peak height (Fig.\,\ref{Fig:QP_spectral_weight}(b)) continuously and smoothly decreases with temperature. Therefore, a second screening mechanism apart from spin screening is active in \RFA~up to at least 300\,K.

Several theoretical studies identify a separation of spin and orbital screening as a defining feature of Hund's metals (see Fig.~\ref{Fig:screening}) \cite{Stadler_2019,deng_2019,horvat2019,Stadler_2021}.
Orbital screening leads to the formation of an orbital singlet and is predicted to set in far above room temperature. The wide cross-over regime is characterized by a bad metallic behavior of incoherent quasiparticles.
At low temperatures, additional spin screening of local moments favors a spin singlet state and leads to the appearance of long-lived quasiparticles.

The characteristic spectral weight changes both at low and at high energies observed in our work provide direct experimental evidence for spin-orbit separation in \RFA. The sudden appearance of the $d_{xy}$ quasiparticle at 90\,K in \RFA~coincides with spin screening. We propose that the continuous changes of the ARPES intensity at low and high energies are in turn the result of orbital screening. This combined action drives the formation of long-lived, renormalized quasiparticles at the Fermi surface and turns a bad metal into a heavy Fermi liquid. We summarize this behavior in Fig.~\ref{Fig:screening}. The spin screening regime is identified by a crossover in the susceptibility, which coincides with a crossover in the resistivity towards a Fermi liquid behavior. Both transport and thermodynamics probe the behavior of the quasiparticles. The corresponding $d_{xz}$ and $d_{xy}$ quasiparticle intensities signify the spectral changes in the two distinct regimes of spin and orbital screening. 

Our study serves as a benchmark for theoretical descriptions of Hund’s metals. Besides separate spin and orbital screening, we observe strong orbital differentiation. Models that include both effects are desirable to obtain a comprehensive picture of correlated multi-orbital systems.


\section*{Methods}

\subsection*{ARPES Experiments}

We synthesized single crystals of \RFA~using growth methods described earlier \cite{Chu_2009}. The samples were prepared and glued onto samples holders for ARPES experiments inside an argon glovebox to minimize air exposure. ARPES experiments were performed at the Stanford Synchrotron Radiation Light Source at beamline 5-2. All samples were cleaved in-situ below 30K. The pressure during cleaving and measurements was below $3\cdot10^{-11}$ torr. \RFA~is known to develop a $\sqrt{2}\times\sqrt{2}$ surface reconstruction, which is common in 122 FeSC \cite{Hoffman_2011,Richard_2018,Pfau_2020}. To remove spectral signatures from the surface bands, we cycled the temperature up to 300\,K and spectra are acquired during cooling (sample 1). Additional analysis was performed on spectra during warm-up (sample 2, Supplementary Fig. S1,S2 \cite{supplement}). We used a photon energy of 60eV with linear vertical polarization.

Quantitative spectral weight measurements as function of temperature are challenging. On one hand, the soft nature of \RFA~crystals leads to uneven sample surfaces after cleaving with only small areas of homogeneous, flat surfaces. Therefore, we use a small beam spot of approximately 50$\mu$m diameter and map the photoemission signal on the whole sample surface. This optimization procedure is repeated after each temperature change to correct for thermal expansion of the sample manipulator and ensure that we probe the same sample spot each time. On the other hand, drifts of the photon flux leads to overall changes in the photoemission intensity. Therefore, we obtain spectra down to 8.5 eV binding energy, which is beyond the As $4p$ bands and Fe $3d$ atomic excitations. No temperature-induced changes are expected in this energy range. Our careful aligning procedures lead to an overall variation of intensity in this region of only 4\%. We normalize all our spectra to the intensity between 7\,eV and 8\,eV. Afterwards, the data fall on top of each other between 6.5\,eV and 8.5\,eV. All intensities are corrected for detector non-linearities.

\subsection*{DFT+DMFT Calculations}

For the DFT+DMFT calculations we performed fully charge self-consistent calculations based on Wien2K v23.2\cite{Wien2k_a,Wien2k_b} in the Generalized-gradient approximation\cite{GGA} and a projection on the subspace of the correlated Fe $3d$ orbitals\cite{Aichhorn2009,Ferber2014} in the window $[-6,13.6]$eV (As $p$, Fe $d$ and higher energy unoccupied states). For solving the impurity model we used continuous-time quantum Monte Carlo method in the hybridization expansion, using the segment picture\cite{Werner2006,Wallerberger2018,Bauer_2011}. We used interaction parameters of $U_{avg}=4$eV, $J_{avg}=0.8$eV, representative for the iron-based pnictides\cite{Roekeghem2016}, in the definition of Slater integrals\cite{SlaterIntegrals}, and the nominal double counting correction\cite{nominalDC1,nominalDC2} with $N=5.5$ nominal filling. 
The DMFT calculations were done at a temperature of $T=96K$, unless indicated otherwise. Stochastic analytical continuation\cite{beach2004} was used to obtain real frequency data.

\subsection*{Atomic Model}

For the simplified impurity model we calculated the spectral function using exact diagonalization of a degenerate five orbital atom system, with each orbital coupled to one non-interacting bath site, and the same form of the local Coulomb interaction as for the DMFT calculation (see above). The hopping parameter $t$ corresponds to the hybridization amplitude to the bath of each orbital (identical for all orbitals). The impurity local chemical potential (energy shift between impurity and bath sites) was adjusted for $N=6$ electron average impurity filling.


 \begin{acknowledgments}
We are very grateful for valuable discussions with Jan von Delft, Rudi Hackl, Yu He, Patrick Kirchmann, Seung-Sup Lee, Brian Moritz. This work is supported by the U.S. Department of Energy, Office of Science, Office of Basic Energy Sciences, Materials Sciences and Engineering Division, under Award Number DE-SC0024135. RV acknowledges support from the Deutsche Forschungsgemeinschaft (DFG, German Research Foundation) -- CRC 1487, “Iron, upgraded!” -- project number 443703006
as well as QUAST-FOR5249 -- project number 449872909. MH and DL acknowledge the support of the U.S. Department of Energy, Office of Science, Office of Basic Energy Sciences, Division of Material Sciences and Engineering, under Contract No. DE-AC02-76SF00515. This research used resources of the Advanced Light Source, which is a DOE Office of Science User Facility under contract no. DE-AC02-05CH11231. Use of the Stanford Synchrotron Radiation Lightsource, SLAC National Accelerator Laboratory, is supported by the U.S. Department of Energy, Office of Science, Office of Basic Energy Sciences under Contract No. DE-AC02-76SF00515.
\end{acknowledgments}


\bibliography{main}

\end{document}


\title{Supplementary Information:\\Observation of Two Cascading Screening Processes in an Iron-based Superconductor}


\author{Ming-Hua Chang}
\email{These authors contributed equally.}
\affiliation{Department of Physics, The Pennsylvania State University, University Park, Pennsylvania, 16802 USA}
\author{Steffen Backes}
\email{These authors contributed equally.}
\affiliation{RIKEN iTHEMS,  Wako, Saitama 351-0198, Japan; Center for Emergent Matter Science, RIKEN, Wako, Saitama 351-0198, Japan}
\author{Donghui Lu}
\affiliation{Stanford Synchrotron Radiation Lightsource, SLAC National Accelerator Laboratory, Menlo Park, California, 94025 USA}
\author{Nicolas Gauthier}
\affiliation{Institut National de la Recherche Scientifique – Energie Matériaux Télécommunications, Varennes, QC J3X 1S2 Canada}
\author{Makoto Hashimoto}
\affiliation{Stanford Synchrotron Radiation Lightsource, SLAC National Accelerator Laboratory, Menlo Park, California, 94025 USA}
\author{Guan-Yu Chen}
\affiliation{Center for Superconducting Physics and Materials, National Laboratory of Solid State Microstructures and Department of Physics, Nanjing University, Nanjing 210093, China}
\author{Hai-Hu Wen}
\affiliation{Center for Superconducting Physics and Materials, National Laboratory of Solid State Microstructures and Department of Physics, Nanjing University, Nanjing 210093, China}
\author{Sung-Kwan Mo}
\affiliation{Advanced Light Source, Lawrence Berkeley National Laboratory, Berkeley, California, 94720 USA}
\author{Zhi-Xun Shen}
\affiliation{Stanford Institute of Materials and Energy Sciences, SLAC National Accelerator Laboratory, Menlo Park, California, 94025 USA}
\affiliation{Department of Physics, Stanford University, Stanford, California, 94305 USA}
\affiliation{Geballe Laboratory for Advanced Materials, Department of Applied Physics, Stanford University, Stanford, California, 94305 USA}
\author{Roser Valent\'\i}
\email{valenti@itp.uni-frankfurt.de}
\affiliation{Institut f\"{u}r Theoretische Physik, Goethe-Universit\"{a}t Frankfurt, Max-von-Laue-Str.~1, 60438 Frankfurt am Main, Germany}
\author{Heike Pfau}
\email{heike.pfau@psu.edu}
\affiliation{Department of Physics, The Pennsylvania State University, University Park, Pennsylvania, 16802 USA}

\maketitle

\begin{figure*}
\includegraphics[width=\textwidth]{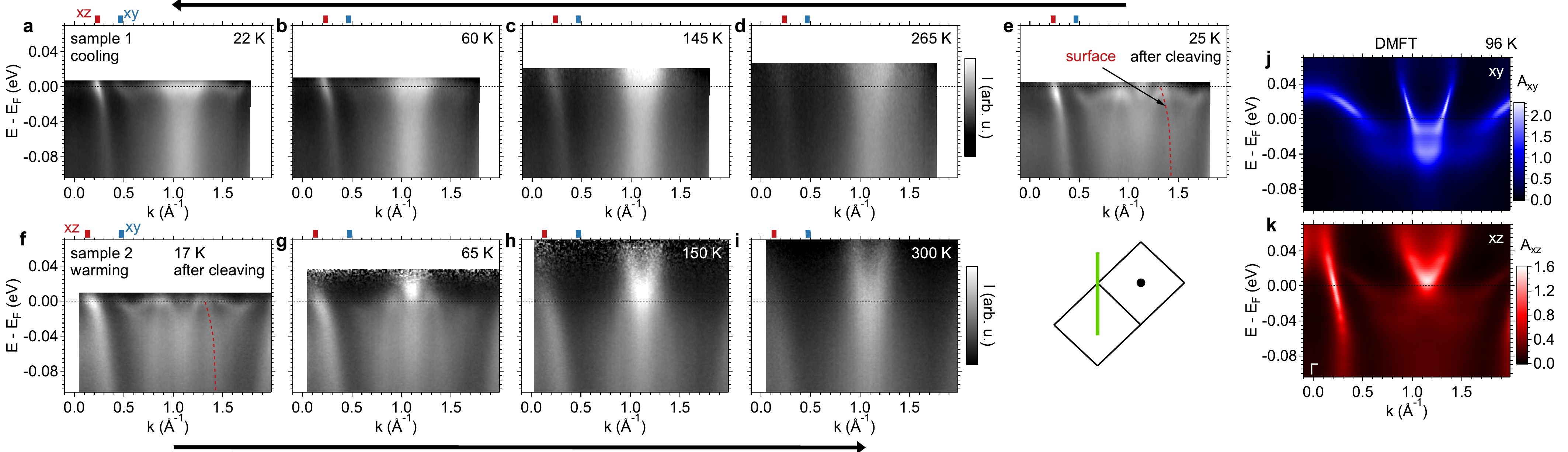}
\caption{
Temperature dependence of the low-energy electronic structure. {\bf{a-d}}: ARPES spectra on sample 1 for selected temperatures. The green line
in the inset sketches the momentum cut through the Brillouin zones. The dot indicates normal emission geometry. The spectra are divided by a Fermi-Dirac distribution and the intensity scale is the same for all spectra. The spectra were taken during cooling after warming up to room temperature. {\bf{e}} shows a spectrum right after cleaving. We highlight one branch of a surface band, which is due to the $\sqrt{2}\times\sqrt{2}$ surface reconstruction, with a dashed line. The arrow on top of the spectra indicates the order in which the spectra were taken. {\bf{f-i}}: ARPES spectra on sample 2 for selected temperatures and along the same momentum cut as in {\bf{a-e}}. The spectra were taken during warming without temperature cycling, see arrow below. A better controlled background intensity in the experiment on sample 2 allows us to retrieve a larger energy window above $E_F$ after division by a Fermi-Dirac distribution. The red and blue squares in {\bf{a-i}} indicate the momentum integration range for the EDC analysis shown in Fig.~\ref{Fig:QP_spectral_weight_s2} and Fig.~2 in the main text. They are labeled with the corresponding orbital character. {\bf{j}}: Spectral function calculated by DFT+DMFT and projected onto the $d_{xy}$ orbital. {\bf{k}}: Same as in {\bf{j}} but projected onto the $d_{xz}$ orbital.
}
\label{Fig:spectra}
\end{figure*}

\begin{figure*}
\includegraphics[width={0.8\textwidth}]{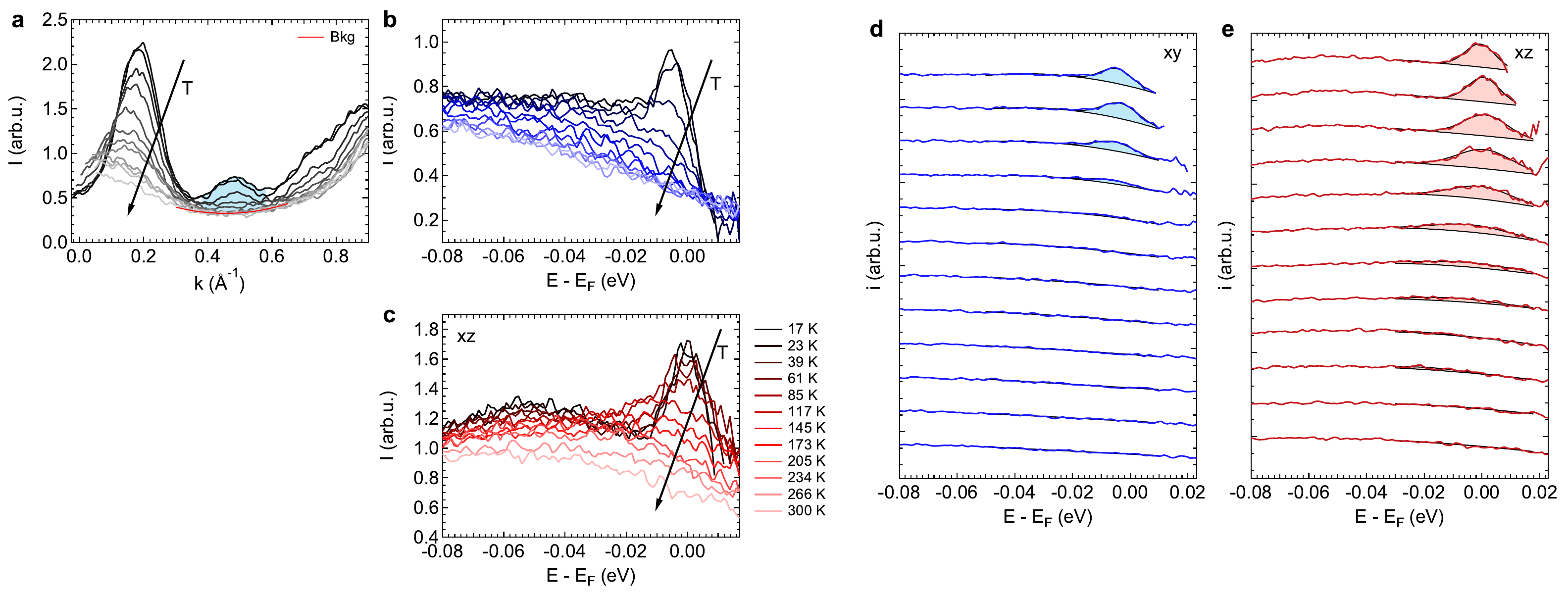}
\caption{
MDC and EDC analysis of sample 2. {\bf{a}}: MDCs integrated over $\pm5$\,meV around $E_F$. The size of the shaded area is plotted in Fig.~2{\bf{b}} of the main text. {\bf{b,c}}: EDCs at $k_F$ of the $d_{xz}$ and $d_{xy}$ hole bands as indicated by the boxes above the spectra in \ref{Fig:spectra}. Fit of EDcs with a sum of a Gaussian and a quadratic background. The shaded area is plotted in Fig.~2{\bf{e}} in the main text.
}
\label{Fig:QP_spectral_weight_s2}
\end{figure*}

\begin{figure*}
\includegraphics[width={0.8\textwidth}]{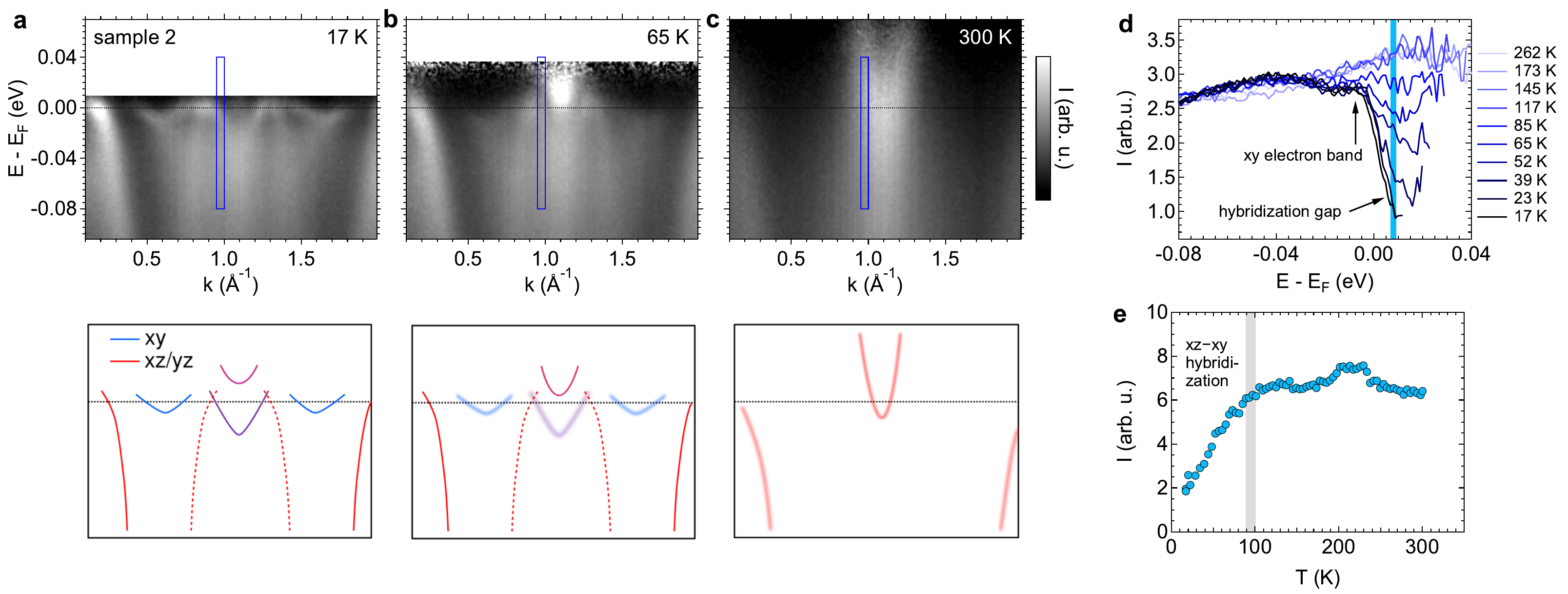}
\caption{
Temperature dependence of the electron bands at the Brillouin zone corner. (a-c) Spectra of sample 2, reproduced from Fig.~\ref{Fig:spectra}. (d) EDCs integrated in the momentum range marked by the boxes ion (a-c). (e) Spectral weight obtained from the EDCs in (d) integrated within the energy range marked within the blue shaded area. A clear drop can be identified at around 90\,K. This is the same temperature, where the $d_{xy}$ hole band disappears. The sketches of the band structure in the three bottom left panels sketch our interpretation of the temperature dependence seen at the Brillouin corner: At low temperatures, the $d_{xy}$ and $d_{xz}$ electron bands hybridize as is commonly observed in FeSC \cite{yi_2019_nematic}. This leads to a suppression of spectral weight inside the hybridization gap as seen in the EDCs in (d). When the $d_{xy}$ orbital disappears, also the hybridization gap vanishes and the intensity in (e) becomes constant.
}
\label{Fig:electron_bands}
\end{figure*}

\begin{figure*}
\includegraphics[width={0.6\textwidth}]{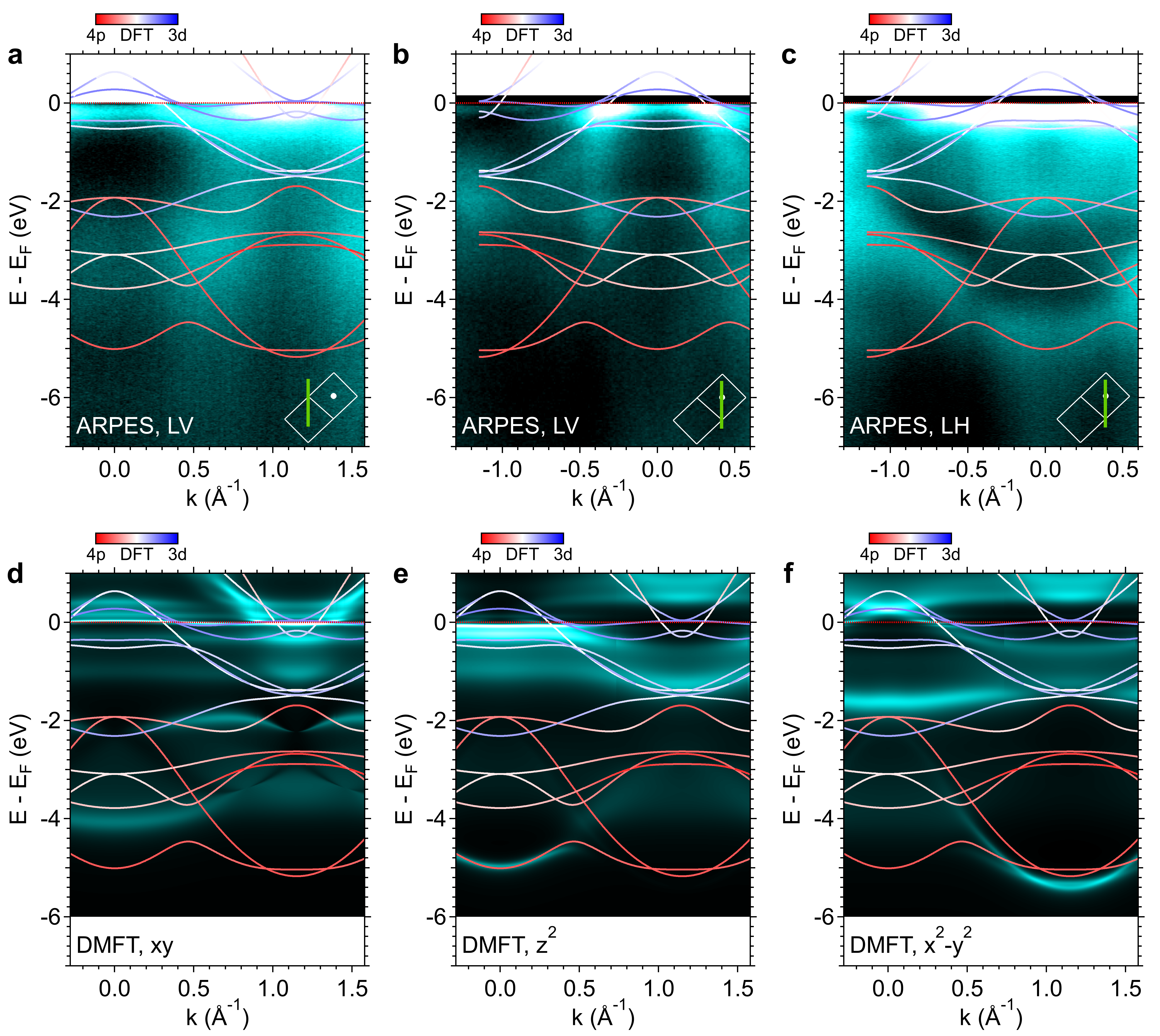}
\caption{
ARPES and DMFT on \RFA~compared with DFT. DFT dispersions are scaled overall by 1.15 to match the ARPES data for $4p$ bands in {\bf{c}} and the DMFT data in {\bf{e,f}}. Insets in (a-c) show the momentum cut through the Brillouin zone for the ARPES measurements. LV and LH denote linear vertically and linear horizontally polarized light. DMFT spectral functions in (d-f) are projected onto different orbital characters as indicated in each panel.
}
\label{Fig:Rb122_DFT}
\end{figure*}

\begin{figure*}
    \centering
    \includegraphics[width=0.8\textwidth]{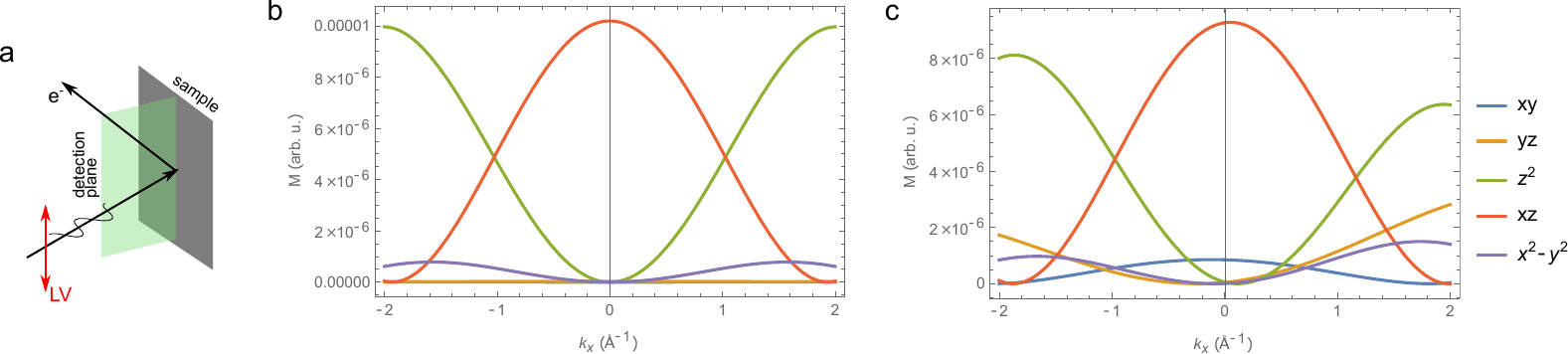}
    \caption{Photoemission dipole matrix elements of the Fe $3d$ orbitals. (a) Experimental geometry. (b) Matrix elements for the spectrum in Fig.~3f of the mina text. (c) Matrix elements for all other ARPES spectra. The calculations were performed in the length gauge using the approximation of a free electron removal state and Fe$3d$ hydrogen-like wave functions as initial states. \cite{goldberg_1978,gadzuk_1975}. The results are in agreement with photoemission studies on various FeSCs, see \eg~\cite{yi_2019_nematic,pfau2019detailed}.
    }
    \label{fig:matrix_elements}
\end{figure*}

\FloatBarrier

\section*{Supplementary Note 1. Temperature-dependent weight from $p-d$ hybridization }

\begin{figure}[h]
    \centering
    \includegraphics[width=1\columnwidth]{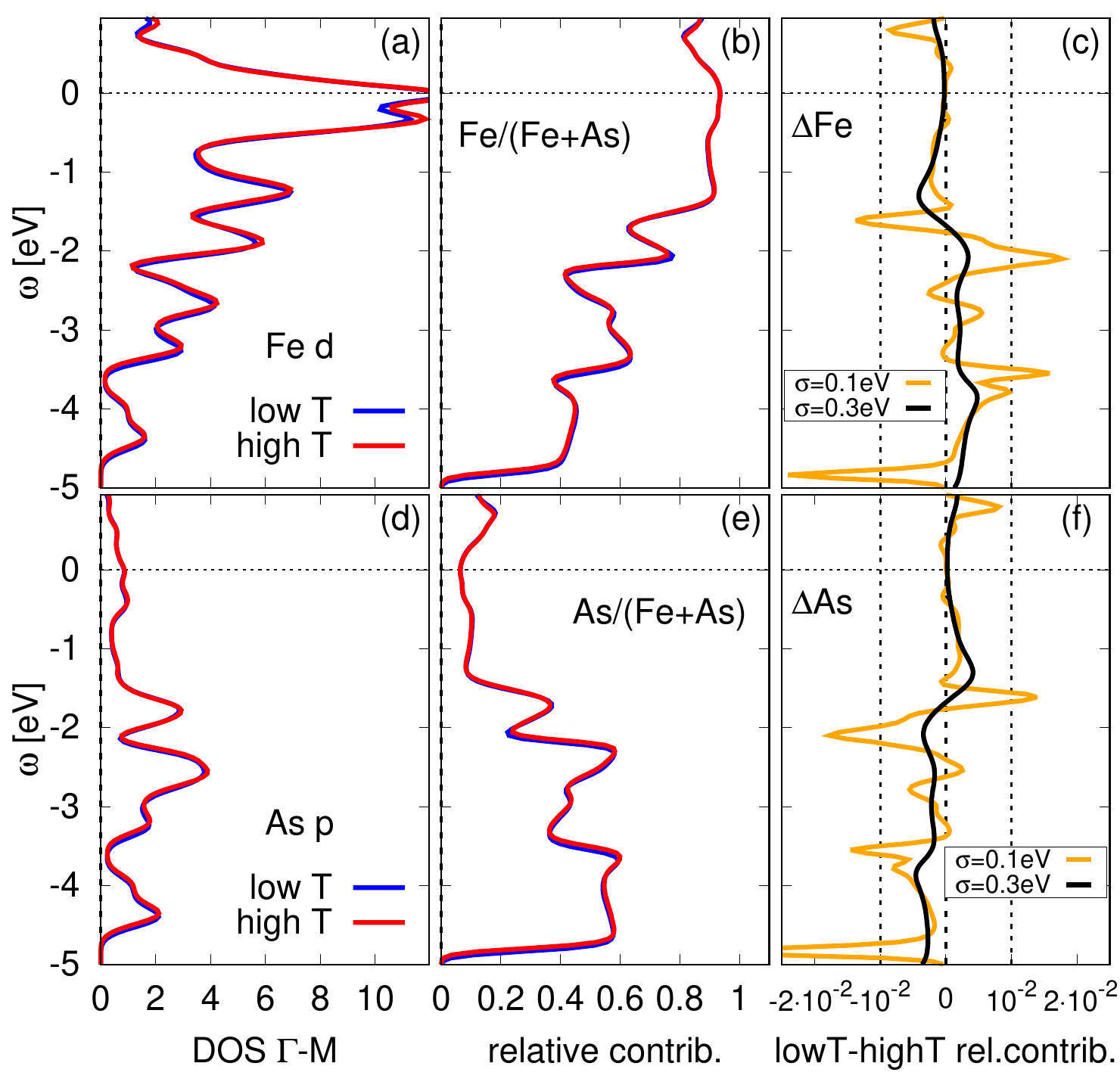}
    \caption{(a) The Fe $3d$ projected spectral weight (DOS), integrated along the $\Gamma-M$ path of the Brillouin zone, as obtained from DFT, comparing the low (blue line) and high temperature (red line) structure.
    (b) The relative contribution of Fe $3d$ to the total spectral weight for the two different temperatures.   
    (c) Difference between the relative contribution at low and high temperature for two values of broadening $\sigma$. We applied a Gaussian broadening to calculate the DOS, and (a) and (b) correspond to $\sigma=0.1$~eV. (d-f) Corresponding quantities for the As $4p$ orbitals.
    }
    \label{fig:lattice_expansion_spectral_weight}
\end{figure}

To investigate the change in Fe $d$-As $p$ hybridization due to lattice expansion, we performed DFT calculations for two different structures: 1) The room temperature structure \cite{Eilers_2016}, and 2) the same structure, but with in-plane lattice parameters a, b reduced by 0.5\%, which simulates the lattice contraction at low temperature \cite{Wiecki_2021}.
The spectral weight, integrated over the same $\Gamma-M$ path as in the ARPES data is shown in~Fig.\ref{fig:lattice_expansion_spectral_weight}. Figure \ref{fig:lattice_expansion_spectral_weight} (a,d) show the spectral weight $w_{Fe}$ and $w_{As}$ for the two structures projected onto the Fe $3d$ and As $4p$ orbitals, respectively. Fig.~\ref{fig:lattice_expansion_spectral_weight} (b,e) display the relative contribution to the total spectral weight $w_{Fe}/(w_{Fe}+w_{As})$ for the Fe $3d$ orbitals, and accordingly for the As $4p$ orbitals. The weight hardly changes as function of temperature. The major difference is a bandwidth reduction of approximately 1\% at high temperature due to the reduction of hybridization strength as the lattice expands.

Fig.~\ref{fig:lattice_expansion_spectral_weight} (c,f) show the difference of the relative weight between low and high temperature $w(T_{low})-w(T_{high})$. The energy scale at high temperature has been rescaled by 1\% to account for the bandwidth reduction. 
Two different Gaussian broadenings have been applied. Broadening by  $\sigma=0.3$\,eV results in a relatively flat energy distribution that most closely resembles the energy dependence of the ARPES spectral weight change. We observe that $d$ spectral weight is shifted from $(-5,-2)$\,eV towards $(-2,0)$\,eV, with a corresponding opposite shift for the As $4p$ orbitals. This is the expected behavior of an increase in $d-p$ hybridization strength due to the lattice contraction. The difference in weight is rather small and of the order of $10^{-3}$, in accordance with the change in latter parameters.

\FloatBarrier


\bibliography{supplement}